\documentclass[fleqn,twoside,12pt]{article}
\usepackage{ae}
\usepackage{aecompl}
\usepackage{amsmath}
\usepackage{amssymb}
\usepackage{epsfig}
\usepackage{graphicx}
\usepackage{espcrc1}

\begin{document}
\thispagestyle{empty}

\title{Interplay between wetting and phase behavior in binary polymer films and wedges: Monte Carlo simulations and mean field calculations}

\author{Marcus M{\"u}ller\thanks{\tt email: Marcus.Mueller@uni-mainz.de} 
and Kurt Binder
\\
Institut f{\"u}r Physik, WA331, Johannes Gutenberg Universit{\"a}t, D55099 Mainz, Germany}
       
\maketitle

\begin{abstract}
Confining a binary mixture, one can profoundly alter its miscibility behavior. The qualitative features of
miscibility in confined geometry are rather universal and shared by polymer mixtures as well as small molecules,
but the unmixing transition in the bulk and the wetting transition are typically well separated in polymer blends.
We study the interplay between wetting and miscibility of a symmetric polymer mixture via large
scale Monte Carlo simulations in the framework of the bond fluctuation model and via numerical self--consistent
field calculations. The film surfaces interact with the monomers via short ranged potentials and the wetting
transition of the semi--infinite system is of first order. It can be accurately located in the simulations
by measuring the surface and interface tensions and using Young's equation.

If both surfaces in a film attract the same component, capillary condensation occurs and the critical point is
close to the critical point of the bulk. If surfaces attract different components, an interface localization/delocalization
occurs which gives rise to phase diagrams with two critical points in the vicinity of the pre-wetting critical point
of the semi--infinite system. The crossover between these two types of phase diagrams as a function of the surface field asymmetry is studied.

We investigate the dependence of the phase diagram on the film thickness $\Delta$ for antisymmetric surface fields.
Upon decreasing the film thickness the two critical points approach the symmetry axis of the phase diagram,
and below a certain thickness $\Delta_{\rm tri}$, there remains only a single critical point at symmetric composition.
This corresponds to a second order interface localisation/delocalisation transition even though the wetting transition is of first order.
At a specific film thickness, $\Delta_{\rm tri}$, tricritical behavior is found.

The behavior of antisymmetric films is compared with the phase behavior in an antisymmetric double wedge.
While the former is the analogon of the wetting transition of a planar surface, the latter is the analogon of the  filling behavior of a single wedge.
We present evidence for a second order interface localization/delocalization transition in an antisymmetric double wedge
and relate its unconventional critical behavior to the predictions of Parry {\em et al.\ } (Phys.Rev.Lett {\bf 83} (1999) 5535)
for wedge filling. The critical
behavior differs from the Ising universality class and is characterized by strong anisotropic fluctuations.
\end{abstract}

%

\newpage
\setcounter{page}{1}
%
%
\section{Introduction}
Confining a binary mixture, one can profoundly alter its miscibility
behavior\cite{DIETRICH,EVANS,PARRY,NAKANISHI}. The phase behavior of $AB$ mixtures in
pores, slits and films has attracted abiding interest from both theorists and
experimentalists\cite{GUBBINS,EXP1,EXP2} alike.  We study the interplay between (pre)wetting 
and phase behavior by self-consistent field (SCF) theory\cite{EPL,SCF} and Monte Carlo simulations\cite{WET,MC,REV}. 
Particularly, we focus on situations where surfaces attract different components 
of the mixture.

The qualitative features of the miscibility in confined geometry are rather
universal and shared by polymer mixtures as well as small molecules.  Symmetric
binary polymer blends are, however, particularly well suited to study the
interplay between wetting and miscibility: (i) the wetting transition
temperature typically is much lower than the critical temperature, where
demixing occurs in the bulk\cite{WET}, (ii) Fluctuations can be controlled by
the degree of interdigitation\cite{SCF,MREV}: the more extended the molecule
is, the larger is the number of neighbors it interacts with, and the smaller
the effect of fluctuations. Therefore SCF calculations provide an accurate description
for many properties except for the ultimate vicinity of critical points. The spatial extension of the molecules also sets
the length scale of enrichment layers and 
facilitates experimental investigations. Indeed, wetting transitions have 
been studied in recent experiments\cite{EXP1,EXP2}. (iii) The vapor pressure of polymer films is
vanishingly small, hence effects of evaporation can be neglected. (iv) Polymers
tend not to crystallize easily. Therefore, wetting phenomena might not be
preempted by crystal phases. Likewise there is no roughening transition of the
interface as it occurs in Ising-like models.

Using a coarse-grained polymer\cite{MREV,BFM} model for a $AB$ binary melt we locate the first
order wetting transition, the phase diagram in a symmetric slit pore (symmetric
film)\cite{WET}, the phase diagram in a thin film where the substrate favor the
$A$-component of the mixture with the same strength as the top surface attracts
the $B$-component (antisymmetric film)\cite{SCF,MC}. Then we discuss the phase behavior in a
quadratic pore where two neighboring surfaces favor the $A$-component and the
other two neighboring surfaces favor the $B$-component (antisymmetric double
wedge)\cite{WEDGE,WEDGE2}. We conclude with an outlook. 

%
\section{Model and techniques}
\begin{figure}[thb]
\begin{minipage}{0.5\linewidth}
\epsfig{file=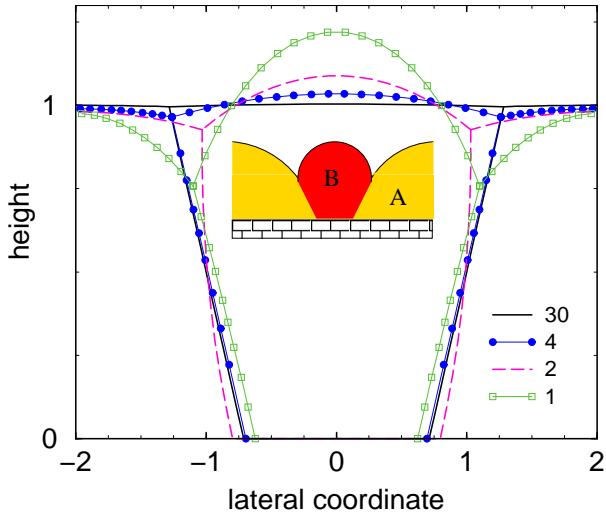,width=\linewidth}
\end{minipage}
\hfill
\begin{minipage}{0.45\linewidth}
\caption{
\label{fig:ill}
Laterally segregated binary film. The shape of interfaces is
obtained by minimizing the effective Hamiltonian \protect{${H}= \gamma_{AB}
L_{AB} +  \gamma_{AS} L_{AS} + \gamma_{BS} L_{BS} + \gamma_{AV} L_{AV} +
\gamma_{BV} L_{BV}$} at fixed volume of the components. $L_{ij}$ denotes the
length of the interface between substances $i$ and $j$, and $\gamma_{ij}$ the
corresponding interface tension. $\gamma_{AV}-\gamma_{BV} =
\gamma_{AS}-\gamma_{BS} = 0.5 \gamma_{AB}$, $\gamma_{BS}= \gamma_{AB}$ and
$\gamma_{AV}/\gamma_{AB}$ as indicated in the key. From \protect\cite{MCOMP}. }
\end{minipage}
\end{figure}

We consider a binary polymer blend. Both species -- $A$ and $B$ -- contain the same
number $N$ of monomers and have the same spatial extension $R_e$. They are confined into
a thin film; the bottom substrate (W) might be a silicon wafer, while the other
surface might be the interface to the vapor (vacuum, V). Depending on the
ratio between the interface tension $\gamma_{AB}$ between the segregated bulk
phases and the surface tension $\gamma_{AV}$, $\gamma_{BV}$ of the components
and the vapor, the upper surface might be rough. The qualitative behavior is
illustrated in Fig.\ref{fig:ill}. 
If the $AB$ interface tension is comparable to the liquid/vapor tension, it
``drags'' the film surface towards the substrate so as to reduce the length of
the $AB$ interface. If the liquid/vapor tension exceeds the $AB$ interface
tension by about two orders of magnitude, however, the surface is almost flat
and the situation is equivalent to a binary mixture between two hard walls a
distance $\Delta$ apart\cite{MCOMP}. In the following we shall restrict ourselves to this
limit $\gamma_{AB} \ll \gamma_{AV}$ or $\gamma_{BV}$.  

In the Monte Carlo simulations we use a computationally efficient,
coarse-grained lattice model. The bond fluctuation model\cite{MREV,BFM}
retains the universal features of polymers -- connectivity, excluded volume of
segments and a thermal interaction which leads to phase separation -- but
ignores details of chemical structure. Effective monomers prevent the
corners of a unit cell of a 3D cubic lattice from double occupancy. We use
chain length $N=32$ and $R_e\approx 17u$.  Monomers along a chain are connected
via bond vectors of length $2, \sqrt{5}, \sqrt{6}, 3$ or $\sqrt{10}$ in units
of the lattice spacing $u$. Different monomers repel each other by a square well
potential of depth $\epsilon$ which comprises the nearest 54 neighbors, like
monomers attract each other.  The strength of the repulsion is proportional to
the Flory--Huggins parameter $\chi=5.3 \epsilon/k_BT$\cite{MREV}. Surfaces are structureless
and impenetrable. They act on monomers in the two nearest layers ($d_{\rm wall}=2$) with strength $\epsilon_{\rm
wall}$.

In the SCF calculations we model the polymers as Gaussian
chains\cite{EPL,SCF,FILM}. The repulsion between different species is quantified
by the Flory-Huggins parameter $\chi$. Short-ranged interactions of strength
$\Lambda_1$ and $\Lambda_2$ attract (repel) the $A$ ($B$) component in the
vicinity of the surfaces.  The total density profile of the film is imposed. It
smoothly decays to zero at the surfaces in a boundary region of width
$0.15R_e$. The blend is assumed to be incompressible. This standard Gaussian
chain model is solved within mean field approximation.

\section{Wetting transition}
\begin{figure}[thb]
\begin{minipage}{0.5\linewidth}
\epsfig{file=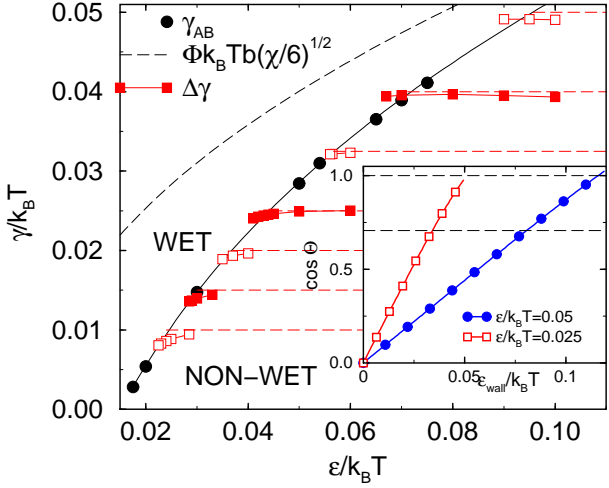,width=\linewidth}
\end{minipage}
\hfill
\begin{minipage}{0.45\linewidth}
\caption{
\label{fig:siwet}
Interface tension $\gamma_{AB}$ and
difference in surface tensions $\Delta \gamma$ as a function of inverse temperature $\epsilon/k_BT$
obtained from simulations.
Approximations for the interface tension $\gamma_{AB}=b\Phi\sqrt{\chi/6}$ and 
$\Delta \gamma = 2\Phi d_{\rm wall} \epsilon_{\rm wall}=\epsilon_{\rm wall}/4$ in the 
strong segregation limit also shown. From \protect\cite{WET}. The inset shows the 
dependence of the contact angle on $\epsilon_{\rm wall}$ for the two temperatures 
investigated in Sec.5.
}
\end{minipage}
\end{figure}
To accurately locate the wetting transition and calculate the contact angle of 
macroscopic $A$-drops we use Young's equation\cite{YOUNG}
$\gamma_{AB}\cos \Theta = \gamma_{WB}- \gamma_{WA}$.
Computationally, this techniques\cite{WET,LV} has distinct advantages for locating first
order wetting transitions: (i)  The interface free energy $\gamma_{AB}$ and the
difference $\Delta\gamma=\gamma_{WB}- \gamma_{WA}$ can be measured accurately
in separate simulations thereby avoiding the need for huge simulation cells to
simulate a thick $A$-layer at the surface in equilibrium with a $B$-rich
bulk.  (ii) Unlike observing the dependence of the thickness of the $A$-layer
on temperature or monomer-surface attraction, one directly measures free
energies. Therefore, we do accurately locate the transition, while the
instability of the $A$-rich layer is located between the transition and the
mean-field wetting spinodal.  (iii) By virtue of the $A \leftrightharpoons B$
symmetry, the difference $\Delta\gamma$ can also be rewritten as the difference
$\Delta\gamma=\gamma_{WB}- \gamma_{-WB}$ of surface tensions of a wall that
attract the $A$-component and a wall that attracts the $B$-component. This
free energy difference can be measured by thermodynamic integration or expanded
ensemble methods\cite{WET}.

The results for our model are presented in Fig.\ref{fig:siwet}. From the
crossing of $\gamma_{AB}(\epsilon)$ and $\Delta \gamma(\epsilon)$ we locate
the wetting transition.  The fact that curves intersect under a finite angle
indicates that the wetting transitions are of first order. As we reduce the
monomer-surface attraction the wetting transition shifts to higher temperatures
$k_BT/\epsilon$ and become weaker. For all transitions studied, however, the
wetting transition is of first order. This is also corroborated by SCF
calculations\cite{SCF,CARMESIN}, where we find first order wetting transitions 
for $T/T_c<0.98$.

If the wetting transition is of first order, then there will be only a small
$A$-rich layer in the non-wet state. By virtue of the structural symmetry of
the molecules, they loose the same amount of entropy as they pack against the
surface. The surface free energy difference $\Delta \gamma$ is mainly enthalpic.
If we assume that the wetting transition is strongly first order, we can
neglect the microscopic enrichment layer at the surface $\Delta \gamma = 2
\epsilon_{\rm wall} d_{\rm wall} \Phi$, where $d_{\rm wall}=2$ denotes the
range of the monomer-surface interaction and $\Phi=1/16$ the monomer number
density.  Using the expression for the interface tension $\gamma_{AB}=\Phi b
\sqrt{\chi/6}$ ($b=3.05$: statistical segment length) in the strong segregation
limit\cite{SSL}, we obtain: 
$\chi_{\rm wet} = 24 \left(\frac{\epsilon_{\rm wet}d_{\rm wet}}{bk_BT} \right)^2$

\begin{figure}[t]
\begin{minipage}{0.5\linewidth}
\epsfig{file=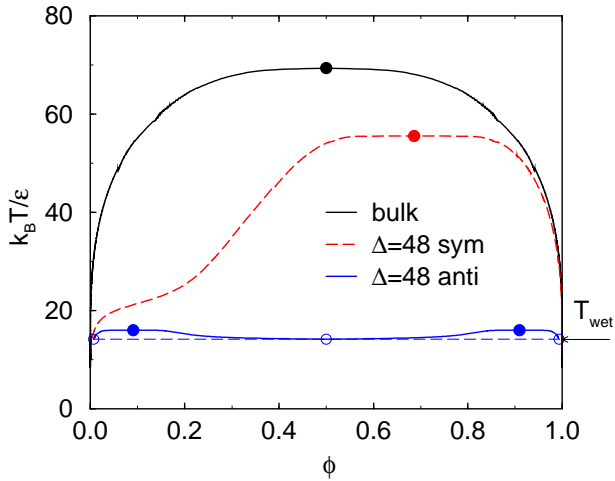,width=\linewidth}
\end{minipage}
\hfill
\begin{minipage}{0.45\linewidth}
\caption{
\label{fig:pdmc}
Phase diagram in terms of composition and temperature for film thickness
$\Delta \approx 2.8 R_e$ as obtained from simulations The arrow 
marks the wetting transition temperature. $\epsilon_{\rm wall}=0.16k_BT$ From \protect\cite{MC}
}
\end{minipage}
\end{figure}
This is in marked contrast to the value of the Flory--Huggins parameter at the
unmixing transition in the bulk, $\chi_c=2/N\sim 1/T_c$. As both the interface tension
$\gamma_{AB}$ and the difference in surface tension $\Delta \gamma$ are chain
length independent, so is the wetting transition temperature. The critical
temperature $T_c$ of phase separation, however, increases linearly with chain length $N$.
Therefore, critical phenomena associated with the bulk unmixing and wetting
phenomena are well separated.

\section{Thin films}
\subsection{Capillary condensation and interface localization/delocalization}

If the mixture is confined into a film the surface interactions modify the phase behavior.
As wetting is associated with the growth of an infinitely large enrichment layer, it is
rounded-off in a thin film\cite{NAKANISHI}. If the wetting transition is of first order,
there will be a pre-wetting transition\cite{DIETRICH}: a coexistence between a thin and a thick (but microscopic)
enrichment layer at a chemical potential which differs from the value at coexistence in the 
bulk. As pre-wetting transitions involve only enrichment layers of finite thickness, they might 
give rise to transitions in thin films.

First we regard a film with symmetric surfaces\cite{WET,REV}, i.e.\ both surfaces attract the $A$ component.
The phase diagram as obtained from the simulations is presented in Fig. \ref{fig:pdmc}.
Compared to the phase behavior in the bulk, the critical point is shifted to
lower temperatures and larger composition of the species attracted by the
surfaces. Moreover, the binodal in the vicinity of the critical points exhibit
2D Ising critical behavior in contrast to the 3D Ising behavior of the bulk
unmixing transition.  

Note the pronounced distortion of the $B$-rich binodal in the vicinity of the
wetting transition. In the $B$-rich phase there are $A$-rich layers at the
surfaces and the $B$ component prevails in the middle of the film. In the
vicinity of the wetting transition the thickness of the $A$-enrichment layers
grows as we increase the temperature.  If we increased the film thickness this
distortion would  evolve into an additional two phase region\cite{WET,TRIPLE}, 
corresponding to a $B$-rich phase with thin and thick $A$-layers at the surface. 
This two phase region would correspond to the pre-wetting coexistence and it 
would join the $B$-rich binodal in a triple point.

\begin{figure}[t]
\epsfig{file=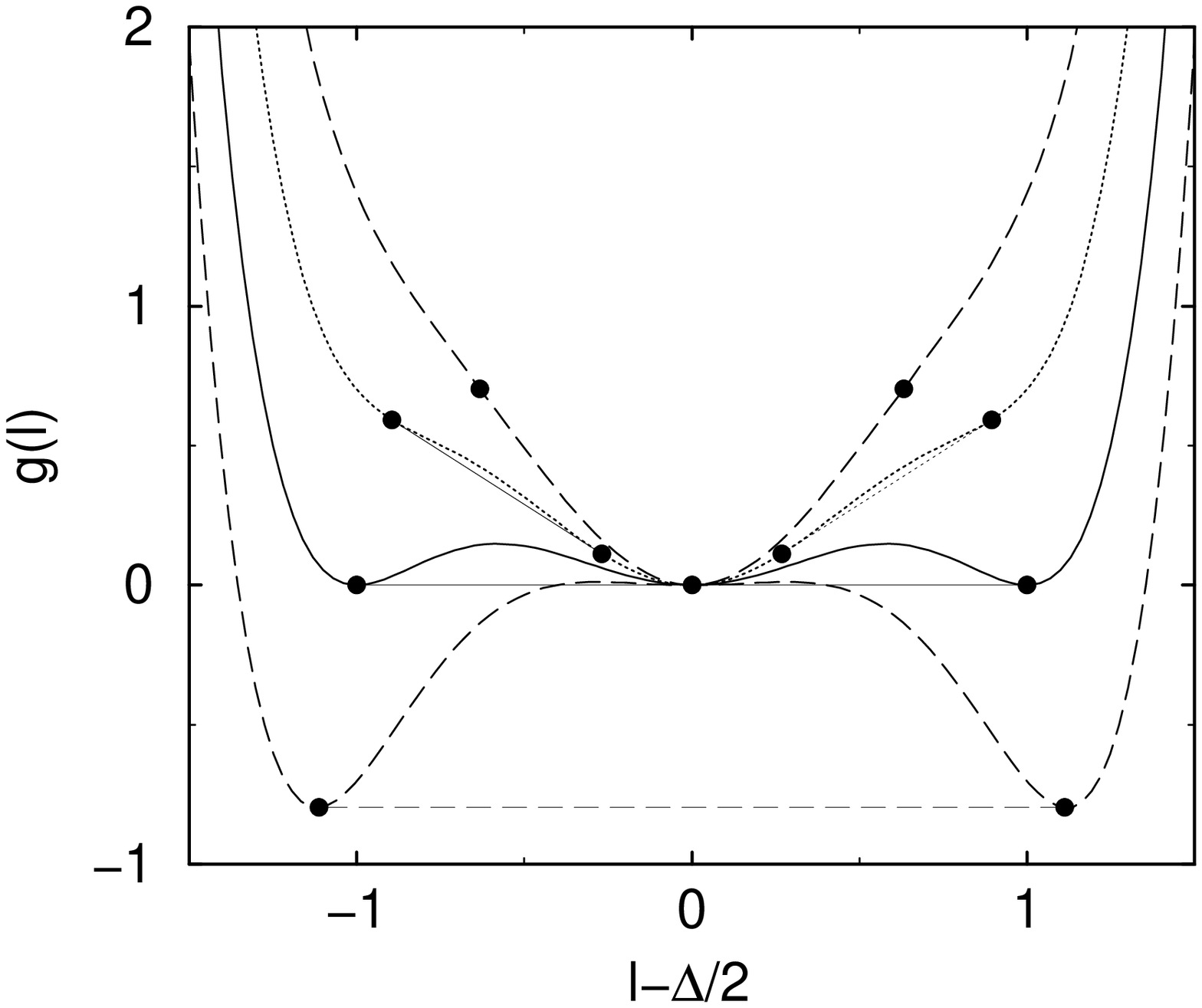,width=0.39\linewidth}
\epsfig{file=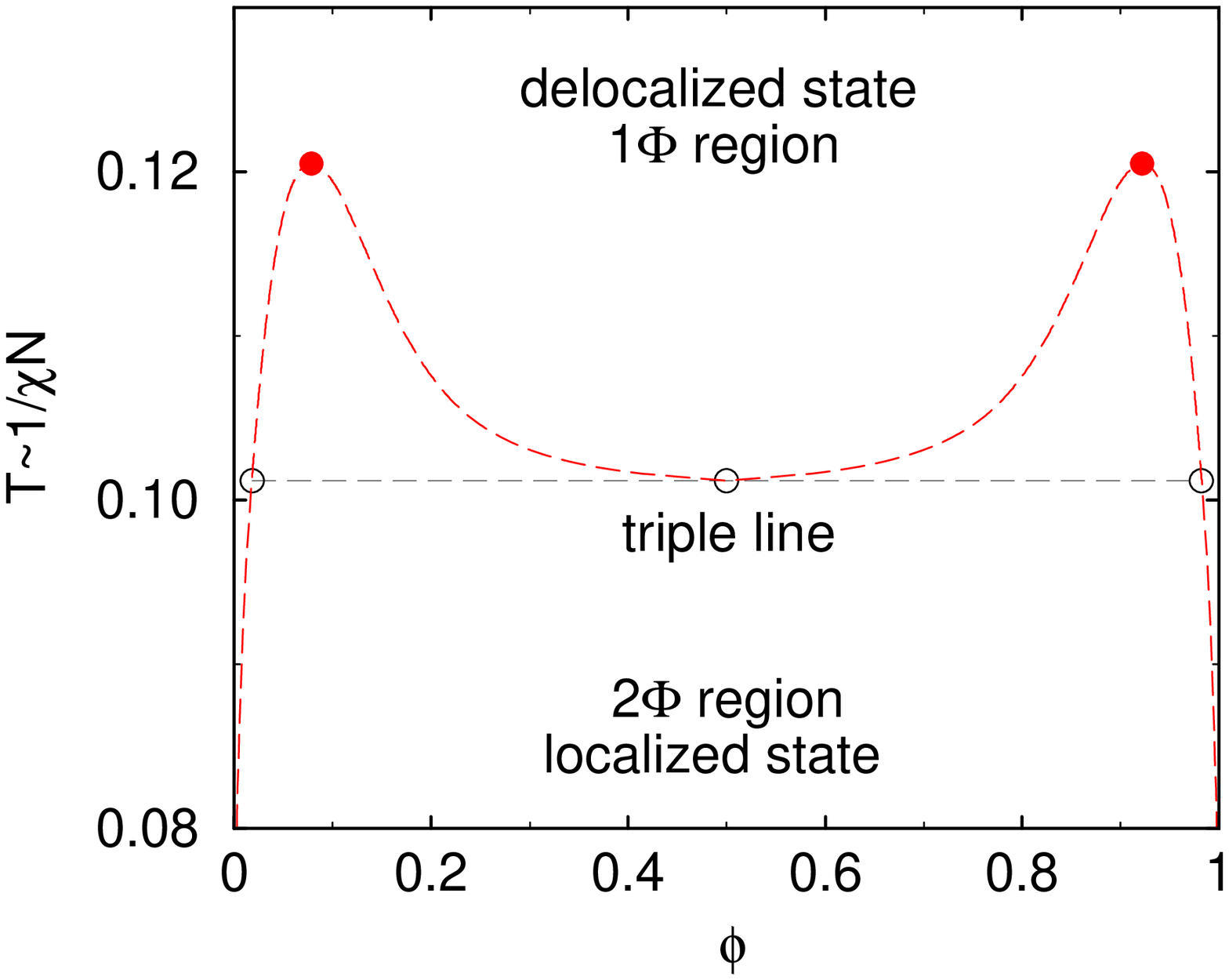,width=0.41\linewidth} \hspace*{2mm}
\epsfig{file=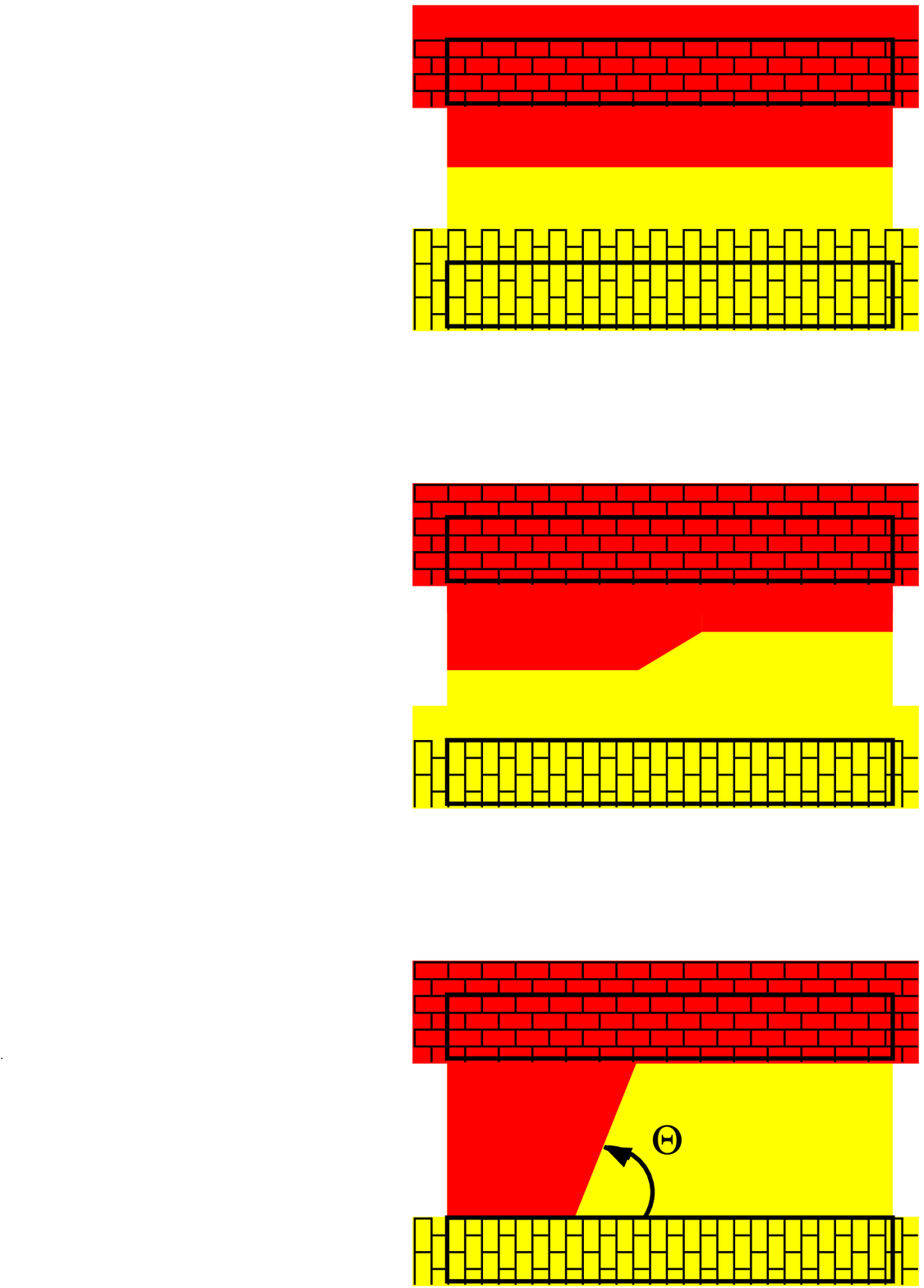,width=0.1\linewidth} \vspace*{-10mm}
\caption{
\label{fig:ginz}
({\bf a}) Schematic temperature dependence of the effective interface potential in a film with
          antisymmetric surfaces. The temperatures correspond to $T<T_{\rm trip}$, $T_{\rm trip}$,
          $T_{\rm trip}< T <T_c^{\rm film}$ and $T_c^{\rm film}$.
({\bf b}) Phase diagram of a mixture.
({\bf c}) Sketches of typical configurations for $T>T_c^{\rm film}$ (upper panel),
          $T_{\rm trip}< T <T_c^{\rm film}$ in the miscibility gap (middle panel)
          and  $T<T_{\rm trip}$ (lower panel). From \protect\cite{REV}
}
\end{figure}
The phase diagram of an antisymmetric film is also presented in
Fig.\ref{fig:pdmc}.  In this case one surface attracts the $A$-component with
exactly the same strength than the other surface the $B$-component. The phase
diagram contains two critical points and a triple line\cite{EPL,SCF,MC}. Around 
the critical temperature of the bulk, enrichment layers gradually form at the surfaces and
stabilize an $AB$ interface that runs parallel to the surfaces. At the
interface localization/delocalization transition\cite{BROCHARD,PE,BINDER} 
this $AB$ interface becomes bound to one of the surfaces. In the case of a first order interface
localization/delocalization transition this corresponds to a triple point of
the phase diagram: an $A$-rich phase , a $B$-rich phase and a phase with
symmetric composition coexist. 

The behavior can be analyzed qualitatively by looking at the interface
potential $g(l)$ which describes the interaction between an $AB$ interface and
a single surface. If the film is thick enough, the interface potential can be
constructed as a superposition of the interface potentials emerging from each
surface. The qualitative behavior in the vicinity of a first order wetting
transition is depicted in Fig.\ref{fig:ginz}({\bf a}).  Using a double-tangent
construction we can construct the phase behavior in a thin film.  At low
temperatures there coexist an $A$-rich phase and a $B$-rich phase, in which the
$AB$ interface is localized at the surface. Upon increasing the temperature,
one encounters the triple point. This triple point is the thin film analogon of
the first order wetting transition. As the film thickness increases the triple
temperature converges towards the wetting transition temperature of the
semi-infinite system. Above the triple temperature there are two phase
coexistence regions, which correspond to thin and thick enrichment layers at the
surfaces. This is the analogon of the pre-wetting transition in a thin film.

\subsection{The tricritical interface localization/delocalization transition}
If we reduce the film thickness, the interactions emerging from each surface
interfere.  The phenomenological considerations\cite{MC} explain that this leads to a
second order interface localization/delocalization transition at small film
thicknesses. Both regimes are separated by a tricritical transition.  
\begin{figure}[t]
\begin{minipage}{0.5\linewidth}
\epsfig{file=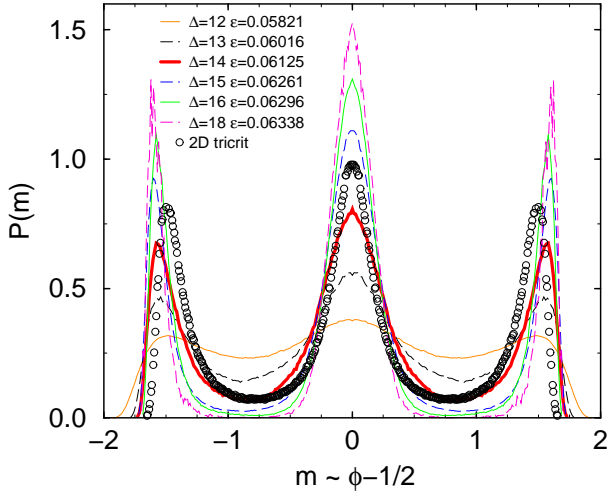,width=\linewidth}
\end{minipage}
\hfill
\begin{minipage}{0.45\linewidth}
\caption{
\label{fig:fsstri}
Probability distribution for various film thicknesses as indicated in the key scaled to unit norm and variance. The lateral system size is $L=96$.
The universal distribution of the 2D tricritical universality class\protect\cite{NIGEL} is presented by circles. From \protect{\cite{MC}}
}
\end{minipage}
\end{figure}
The scaled distribution functions prove convenient to locate the tricritical thickness accurately. To this end we have
adjusted the temperature such that the central peak of the probability distribution of the order parameter $m \sim \phi_A-\phi_B$ is a factor 1.2 higher than the outer peaks. This corresponds to the
behavior of the universal distribution of the 2D tricritical universality class\protect\cite{NIGEL}. The results for various
film thicknesses $\Delta$ (in units of the lattice spacing) are presented in Fig.\ref{fig:fsstri}. For $\Delta<\Delta_{\rm tri}$ the valleys between the three
peaks are too shallow (cf.\ Fig.\ref{fig:fsstri}), while they are too deep for $\Delta>\Delta_{\rm tri}$. In the latter case the transition
is of first order and our estimate tends towards the triple temperature. At $\Delta_{\rm tri}\approx 14 = 0.89 R_e$ the distribution of our
simulations is similar to the universal 2DT distribution, and this has been confirmed for larger lateral system sizes\cite{MC}.
\subsection{Crossover from capillary condensation to interface localization/delocalization}
\begin{figure}[thb]
\epsfig{file=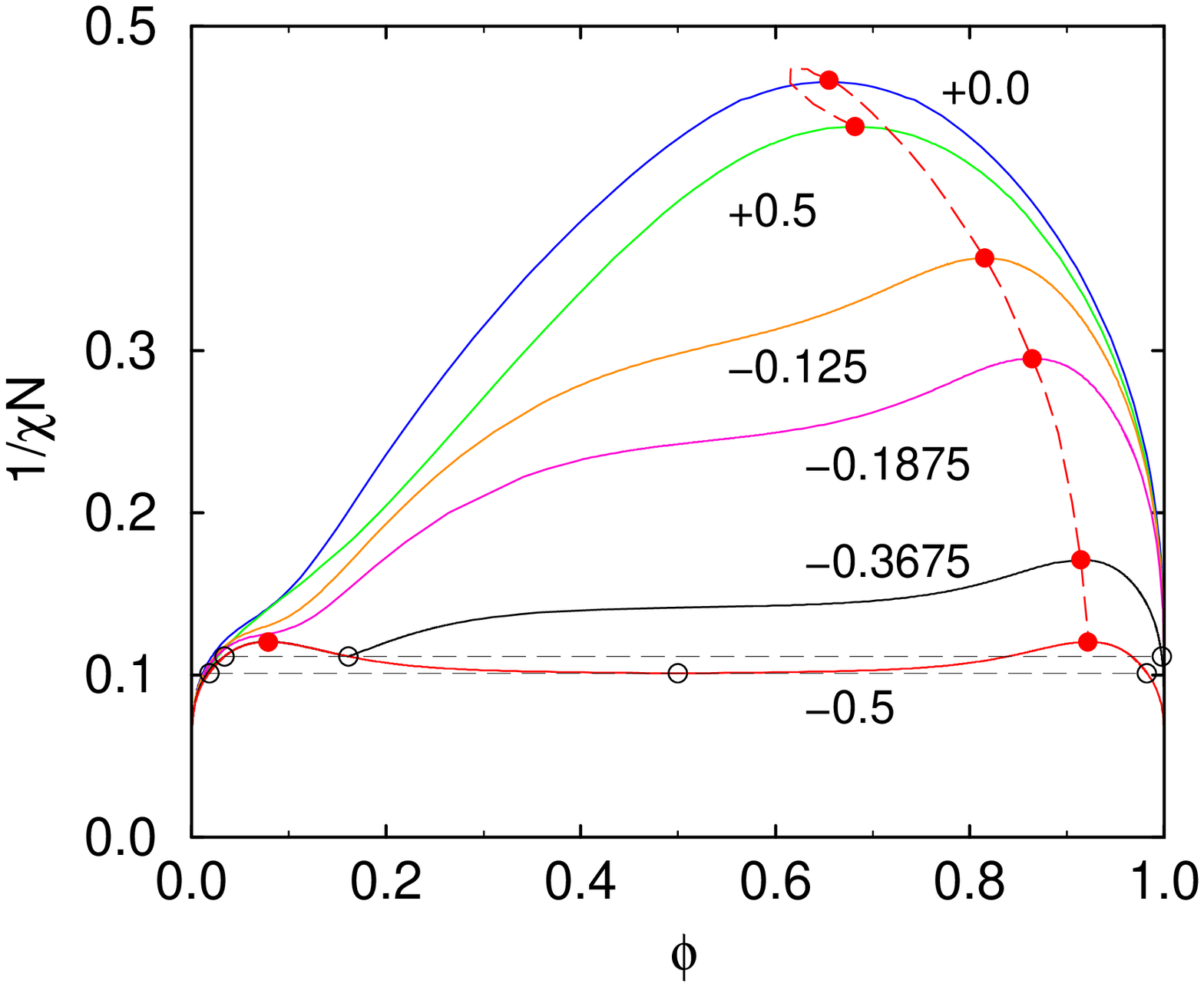,width=0.45\linewidth}
\epsfig{file=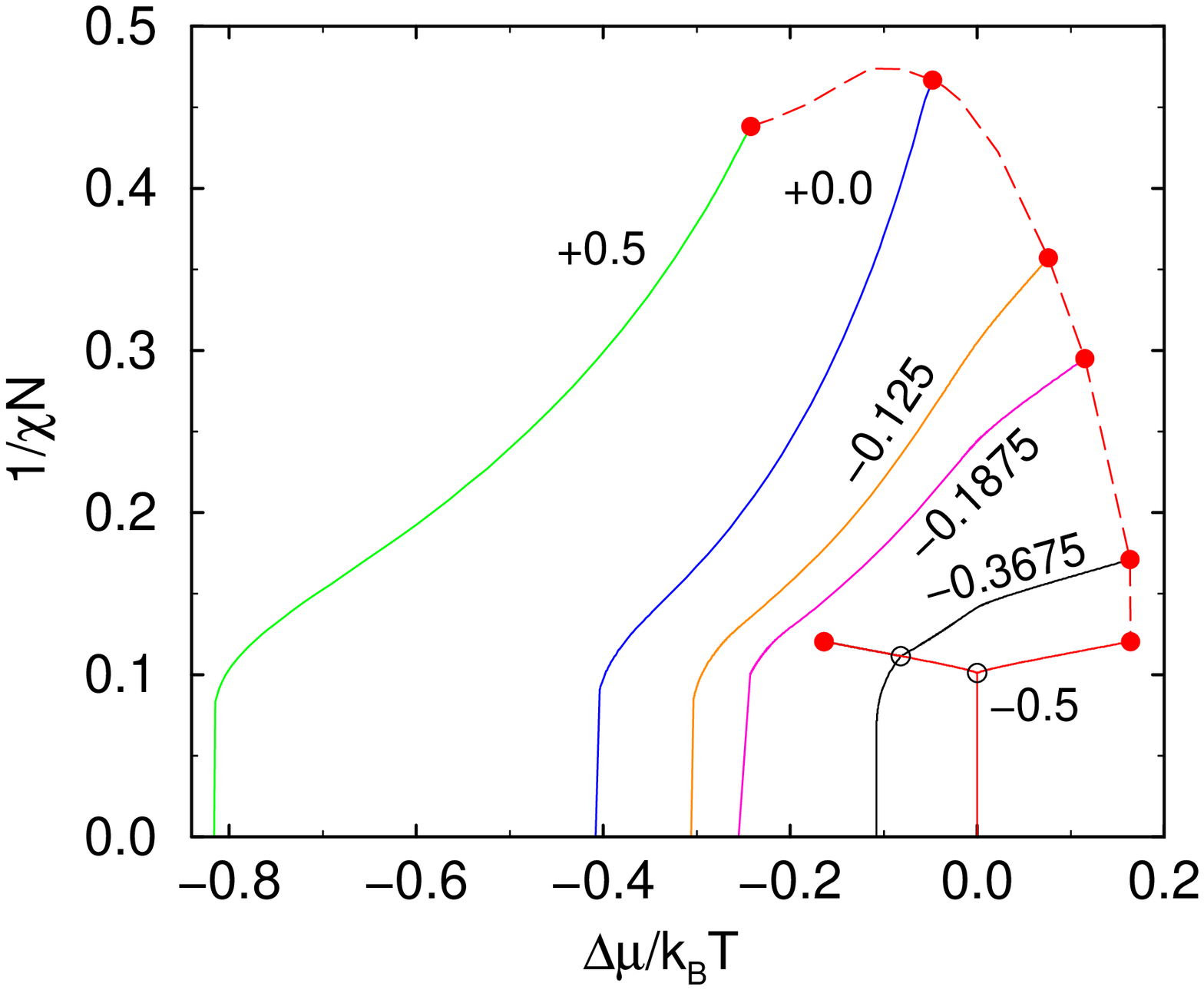,width=0.45\linewidth}\vspace*{-10mm}
\caption{
\label{fig:cross}
({\bf a}) Binodals for $\Delta_0=2.6 R_e$ and surface interaction $\Lambda_1 N=0.5$ obtained from SCF calculations. The surface interaction at the other surface $\Lambda_2 N$ varies as
         indicated in the key.  The dashed curve shows the location of the critical points. Filled circles
         mark critical points, open circles/dashed horizontal lines denote three--\- phase coexistence for
         $\Lambda_2 N=-0.3675$ and $-0.5$.
({\bf b}) Coexistence curves in the $\chi N$-$\Delta\mu$ plane.
         The ``quasi-pre-wetting''
         lines for $\Delta \mu<0$ and $\Lambda_2N=-0.3675$ and $-0.5$ are indistinguishable, because they are
         associated with the pre-wetting behavior of the surface with interaction $\Lambda_1N=+0.5$. From\protect{\cite{EPL}}
	 }
\end{figure}
Realizing strictly (anti)symmetric surface interactions is often difficult in experiments.
Varying the surface interaction $\Lambda_2N$ of the top surface from attracting the
$A$-component to attracting the $B$-component (while the bottom surface always
attracts the $A$-component with fixed strength $\Lambda_1N$) in the SCF calculations, we study the crossover
from capillary condensation for symmetric surfaces to interface
localization/delocalization.  The dependence of the phase diagram on the
surface interactions within the SCF calculations is
presented in Fig.\ref{fig:cross}.  
For symmetric surfaces (capillary condensation) the critical point is shifted
towards lower temperatures\cite{NAKANISHI} similar to the simulation result.  Of course, the
binodals are parabolic in mean field theory independent from
dimensionality. The coexisting phases have almost uniform composition across
the film and differ in their composition. As we reduce the preference of the
top surface for species $B$, the critical point and the critical composition
tend towards their bulk values ($\phi=0.5$, $1/\chi N=0.5$), i.e.\ the critical temperature increases and
the critical composition becomes more symmetric\cite{EPL}.  The coexistence
curve in the $1/\chi N$-$\Delta \mu$ plane approaches the symmetry axis. Upon
making the top surface attracting the other component $B$, we gradually change
the character of the phase transition towards an interface
localization/delocalization transition\cite{BROCHARD,PE}. The critical
temperature passes through a maximum and the critical composition through a
minimum, respectively.  For $\Lambda_2N<0$ (surface attracting the
$B$-component) there are enrichment layers of the $A$-component at the bottom
and the $B$-component at the top, and the two coexisting phases differ in the
location of the $AB$ interface which runs parallel to the surfaces. As the
preferential interaction of the top surface increases, the critical temperature
decreases and the critical composition becomes richer in $A$.  When the
coexistence curve intersects the pre-wetting line of the bottom surface at
$\Delta \mu<0$, a triple point forms at which an $A$-rich phase and two
$B$-rich phases with a thin and a thick $A$-enrichment layer coexist. When
the bottom surface attracts $A$ with exactly the same strength as the top
surface $B$ (antisymmetric surfaces), the phase diagram becomes symmetric. 

For symmetric surface fields the critical point occurs close to the bulk critical
point (and converges towards it in the limit of infinite film thickness) while
the critical points in antisymmetric films are associated with the wetting
transition and converge towards the pre-wetting critical temperature of the
semi-infinite system (if the wetting transition is of first order)
for $\delta \to \infty$. In both cases, however, critical points belong
to the 2D Ising universality class.

\section{Interface localization/delocalization in an antisymmetric double wedge}
\subsection{Background}
\begin{figure}[t]
\begin{minipage}{0.5\linewidth}
\epsfig{file=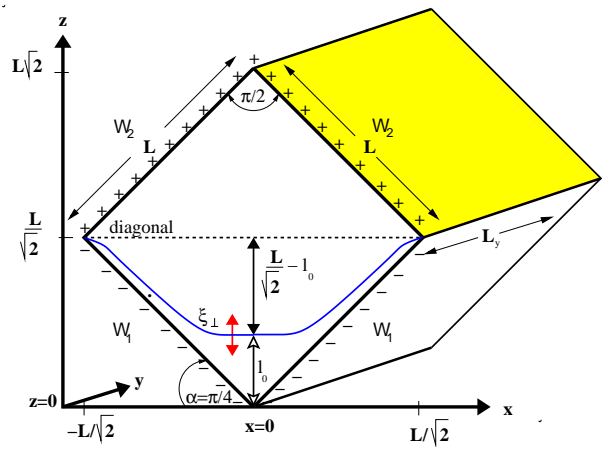,width=\linewidth}
\end{minipage}
\hfill
\begin{minipage}{0.45\linewidth}
\caption{
\label{fig:wedge} Antisymmetric double wedge:
periodic boundary conditions apply along the $y$-direction and there are 4 impenetrable surfaces of size $L \times L_y$.
The bottom ones ($W_1$) attract the $A$-component with strength $\epsilon_{\rm wall}$ and the top ones ($W_2$)
attract the $B$-component. $l_0$ denotes the position of the interface
from one corner. From \protect\cite{WEDGE}.
}
\end{minipage}
\end{figure}
In the following we consider wetting (or rather filling) in a wedge geometry.
Macroscopic considerations show that the wedge will be filled with liquid when
the contact angle $\Theta$ on a planar substrate equals the opening angle
$\alpha$. Intriguingly, Parry and co-workers\cite{PARRYW} predict that the filling of a
wedge is related to the strong fluctuation regime of critical wetting and that
critical filling may even occur if the concomitant wetting transition of the
planar surface is of first order.  Specifically, they predicted the distance $l_0$ of the
$AB$ interface from the bottom of a wedge to diverge as $l_0 \sim (T_f
-T)^{-\beta_s}$ with $\beta_s=1/4$. Correlations along the wedge and in the
other two directions are characterized by diverging correlation lengths $\xi _y
\sim (T_f -T)^{-\nu_y}$ and $\xi_x \sim \xi_\perp \sim (T_f - T)^{- \nu_\perp}$
with exponents $\nu_y=3/4$ and $\nu_\perp=1/4$, respectively.

In the following we study a wedge with opening angle $\alpha=\pi/4$ of the wedge (c.f.\ Fig.\ref{fig:wedge}).
Similar to the study of wetting an antisymmetric geometry is advantageous. Therefore we
stack two wedges which attract different components on top of each other. This
antisymmetric double wedge is a pore with quadratic cross-section of size $L
\times L$. Let $L_y$ denote the length of the wedge (c.f.\ Fig.\ref{fig:wedge}).  
(i) If we used identical surface
fields on all four free surfaces the analog of capillary condensation would
occur in a wedge, i.e., phase coexistence would be shifted away from the
bulk coexistence curve and the wetting layers would be only metastable (with
respect to ``wedge condensation''). (ii) As the wetting layer grew on all four
surfaces in the case of symmetric boundaries, we would need larger system sizes
to reduce interactions between the wetting layers across the wedge.

The phase behavior in such an antisymmetric double wedge geometry has been
studied recently in the framework of an Ising model\cite{WEDGE,WEDGE2}. When the wetting transition
of the planar substrate was of first order, the wedge filling was also found to
be of first order. When the wetting transition was of second order an
unconventional scaling behavior was observed which is characterized by critical
exponents $\alpha=3/4$, $\beta=0$, and $\gamma=5/4$. Those critical exponents
can be related (see below) to the exponents of critical wedge filling, and the
simulations of the Ising model confirm the predictions of Parry and
co-workers\cite{PARRYW}.

In the following we corroborate these findings in the framework of the Ising
model by our polymer simulations. Moreover, we present evidence for the
unconventional second order transition in an antisymmetric double wedge even
though the wetting transition on a planar substrate is of first order. 

\begin{figure}[t]
\begin{minipage}{0.5\linewidth}
\epsfig{file=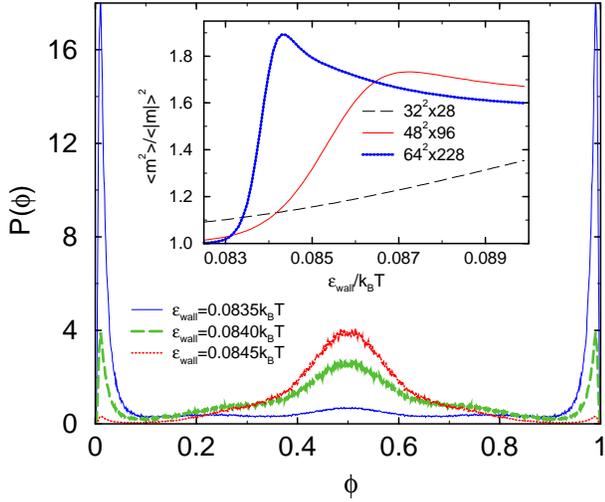,width=\linewidth}
\end{minipage}
\hfill
\begin{minipage}{0.45\linewidth}
\caption{
\label{fig:pfirst} 
\protect{
The probability distribution of the composition at $\epsilon/k_BT=0.05$ and system geometry \protect{$64^2\times 228$}
exhibits a three peak structure, which 
is characteristic of a first order transition. The inset shows the dependence of the cumulant $\langle m^2 \rangle /\langle |m| \rangle^2$
with $m\sim \phi-1/2$ on $\epsilon_{\rm wall}$ for three different system sizes.
}
}
\end{minipage}
\end{figure}
We present preliminary simulation data for two temperatures:
$\epsilon/k_BT=0.025$ ($T/T_c=0.58$) and $\epsilon/k_BT=0.05$ ($T/T_c=0.29$).
At both temperatures the wetting transitions, that occur at appropriate
attractive strength $\epsilon_{\rm wall}$ of planar surfaces, are of first
order (c.f.\ Fig.\ref{fig:siwet}). In the former case it is weak first order
wetting transition, in the latter case it is a strong first order transition.

\subsection{First order transition in an antisymmetric double wedge}
At the lower temperature $\epsilon/k_BT=0.05$, the behavior is similar to a first order interface
localization/delocalization transition.  We consider here only the case $\Delta \mu=0$ 
where phase coexistence in the bulk occurs. This excludes the rather
interesting interplay between pre-wetting and pre-filling behavior studied in
Ref.\cite{REJMER}. At large surface interaction $\epsilon_{\rm wall}>\epsilon_{\rm wall}^{\rm trip,wedge}$
there runs an $AB$ interface along the diagonal which divides the two double
wedges. This corresponds to the delocalized state.  Upon decreasing
$\epsilon_{\rm wall}$ (or decreasing the temperature) the $AB$ interface
becomes localizes in one of the wedges. In this case the composition of the
double wedge is either $A$-rich or $B$-rich and we define as order parameter $m
\equiv \phi_A-\phi_B$.  The two situations are separated by a triple point $\epsilon_{\rm wall}^{\rm trip,wedge}$ at
which the interface can be localized in either of the wedges or delocalized on
the diagonal.  The trimodal probability distribution in the vicinity of the 
tricritical point is presented in Fig.\ref{fig:pfirst}.  In
analogy to the case of antisymmetric films we expect this triple point in a
double wedge to correspond to a first order filling transition. In the inset
we show the cumulant $\langle m^2 \rangle /\langle |m| \rangle^2$. If the transition
was of second order, these cumulants would depend monotonically on $\epsilon_{\rm wall}$ and
would exhibit a common intersection point. This is not at all what we observe, and we conclude
that the interface localization/delocalization transition in the double wedge is of first order at
the lower temperature $\epsilon/k_BT=0.05$.

\subsection{Critical behavior in an antisymmetric double wedge}
\begin{figure}[t]
\epsfig{file=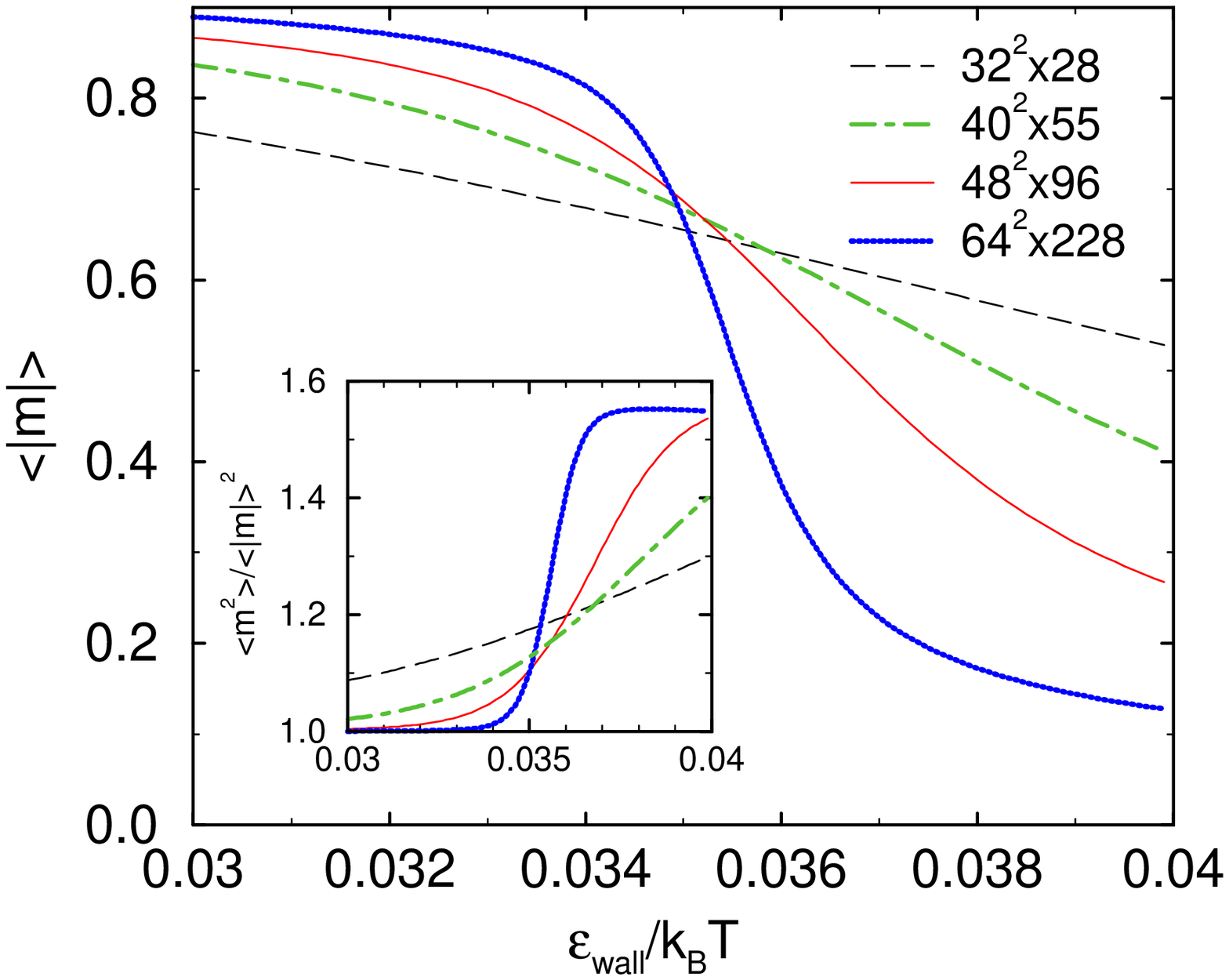,width=0.48\linewidth}
\epsfig{file=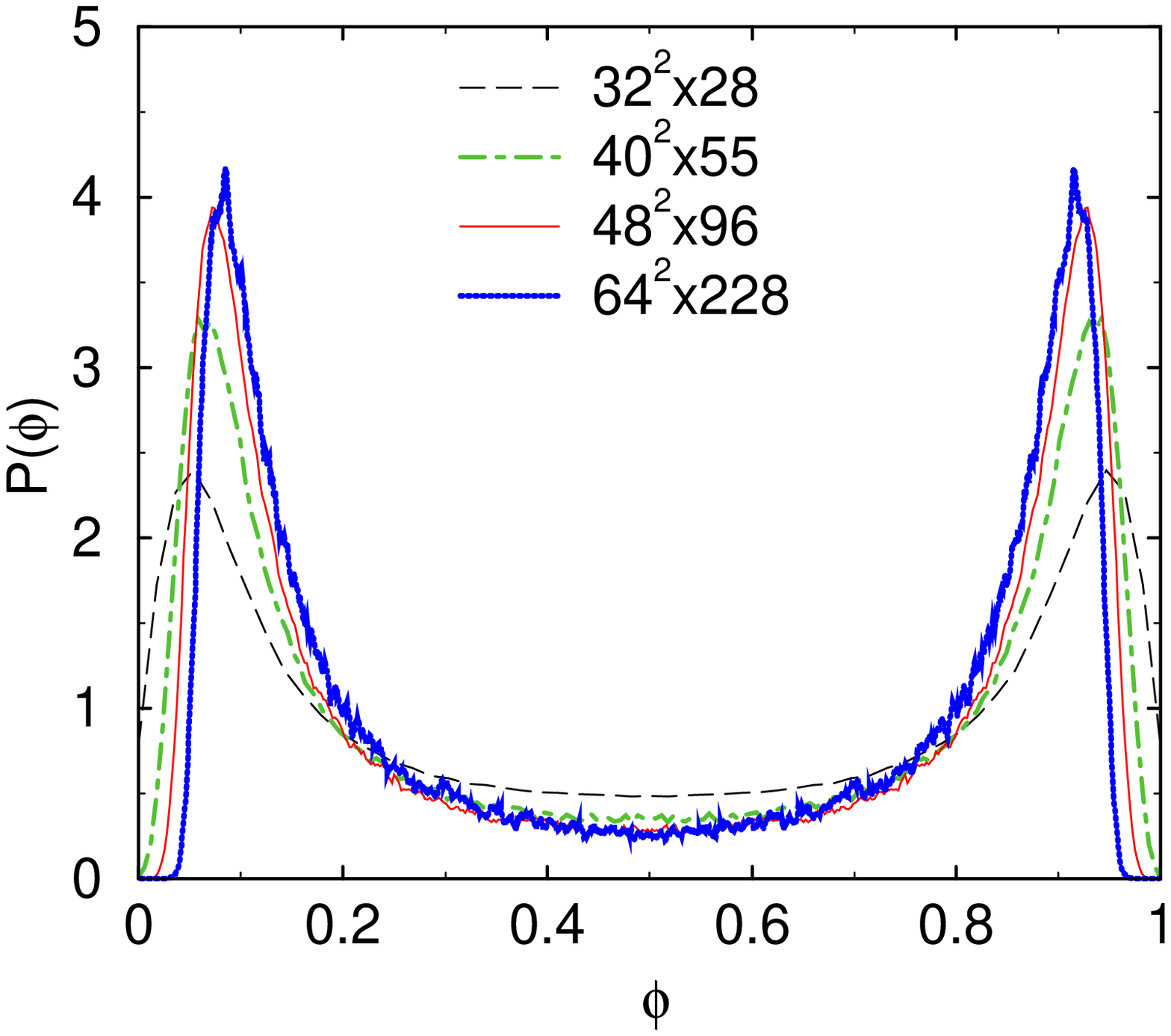,width=0.443\linewidth}\vspace*{-10mm}
\caption{
\label{fig:wcrit}
({\bf a}) Dependence of the absolute order parameter $m \equiv  |\phi_A-\phi_B|$ on the surface interactions $\epsilon_{\rm wall}$
          at $\epsilon/k_BT=0.05$. The insert shows the cumulant.
({\bf b}) Scaling of the probability distribution at $\epsilon_{\rm wall} = 0.035$ and various system sizes.
}
\end{figure}
Even though the wetting transition on a planar surface at $\epsilon/k_BT=0.025$ is of first order,
the behavior at the interface localization/delocalization transition in an antisymmetric double wedge
at high temperature differs from the transition at low temperature. In the inset of Fig.\ref{fig:wcrit} ({\bf a})
we present the dependence of the cumulant on the surface interaction strength for various system sizes.
The cumulants depend monotonically on $\epsilon_{\rm wall}$ and exhibits a common intersection point around 
$\epsilon_{\rm wall}^{\rm crit}\approx 0.035$.  In panel ({\bf b}) we show the probability distribution of 
the composition $\phi$ at this intersection point: the distribution is bimodal and the two largest system sizes 
collapse onto a master curve without any size dependent prefactor. Therefore we conclude that the
interface localization/delocalization transition is of second order.\footnote{The interface localization/delocalization
transition might be of second order in a very thin antisymmetric film (c.f.\ Sec.\ 4.2) even if the wetting transition is
of first order. Therefore still larger system sizes would be desirable to confirm this conclusion. We note however, 
that the thickness of the enrichment layer at the first order wetting transition of the planar substrate ($\epsilon/k_BT=0.0226$,
$\epsilon_{\rm wall}^{\rm wet}/k_BT=0.04$) is only $l_0 \approx 4 \ll 45.2 = L/\sqrt{2}$. Therefore we believe that our conclusion
is not affected by finite size effects.}

Intriguingly there are also marked differences between this second order transition in an antisymmetric double wedge
and the second order transition in a thin film which belongs to the 2D Ising universality class. In the latter case,
only the distribution of the {\em scaled} order parameter $L^{\beta/\nu }m$, where $\beta=1/8$ and $\nu=1$ are the critical
exponents of the order parameter and the correlation length in the 2D Ising universality class, exhibits data collapse
for different system size. Moreover, we present in Fig.\ref{fig:wcrit}({\bf a}) the dependence of the absolute value
of the magnetization in the vicinity of the transition. Curves for different system sizes exhibit a common intersection
point which agrees well with the intersection point of the cumulants. The analogous curves at a Ising-like transition
do not exhibit a common intersection point but monotonously converge towards $\langle |m| \rangle \sim |T-T_c|^\beta$ for
$T<T_c$ and $\langle |m| \rangle\equiv 0$ for $T\geq T_c$ upon increasing the system size.

To relate the critical behavior of the antisymmetric double wedge to the predictions of Parry {\em et al.}\cite{PARRYW},
we regard the distance $l_0$ of the $AB$ interface from the corner of one wedge. 
Similar to an antisymmetric film (c.f.\ Sec.\ 4.1), we assume that we can approximate the distribution in a double wedge
by the superposition of the distributions of single wedges $P_{\rm wedge}(l_0)$ via $P(l_0) \sim P_{\rm wedge}(l_0)+P_{\rm wedge}(\sqrt{2}L- l_0)$
If the two distributions $P_{\rm wedge}(l_0)$ and $P_{\rm wedge}(\sqrt{2}L- l_0)$ do not overlap, the $AB$ interface
will be located in either of the two wedges and the order parameter will not vanish. If the two distributions overlap,
the interface fluctuates around the diagonal and the order parameter will be zero. Right at the transition the two distributions
begin to overlap:
\begin{equation}
\langle l_0 \rangle + \xi_\perp \stackrel{!}{=} \sqrt{2}L- \langle l_0 \rangle - \xi_\perp \qquad \mbox{(interface localization/delocalization in double wedge)}
\end{equation}
where $\langle l_0 \rangle$ denotes the mean height in a single wedge and $\xi_\perp$ its fluctuations.
Importantly, Parry's prediction of $\beta_0 = \nu_\perp$ in wedges (and also corners\cite{CORNER}) means that the
height and its fluctuations are of the same order. They diverge as we approach the critical filling transition.

The height of the interface $l_0$ is related to the order parameter $m$ of the localization/delocalization transition.
Therefore we expect the distribution of the order parameter also to be bimodal. As $l_0 \sim \xi \sim L$ at the transition and
the order parameter is a function of $l_0/L$ the distribution of the order parameter will exhibit two peaks whose positions and 
widths will not depend on the system size. This is exactly what we observe in Fig.\ref{fig:wcrit}({\bf b}).
Using this observation and the standard finite size scaling assumption at a second order phase transitions
\begin{equation}
P(m) \sim L^{\beta/\nu_\perp} \tilde{{\cal P}}(L^{\beta/\nu_\perp} m, L/\xi_\perp,L_y/\xi_y) 
     \sim  L^{\beta/\nu_\perp} {\cal P}(L^{\beta/\nu_\perp} m, L^{1/\nu_\perp}t,\eta)
\end{equation}
where $\tilde{\cal P}$ and ${\cal P}$ are scaling functions, $t=(T-T_f)/T_f$ denotes the relative distance to the filling transition,
and $\eta \equiv L_y/L^{\nu_y/\nu_\perp}=L_y/L^3$ denotes the generalized aspect ratio, we conclude $\beta/\nu_\perp=0$. Due to the
anisotropy of the fluctuations of the interface along the wedge with correlation length $\xi_y$ and perpendicular to the wedge with
correlation length $\xi_\perp$ the generalized aspect ratio appears as a scaling variable. In our simulations we have chosen the
system geometry such that $\eta$ remains approximately constant to ensure that finite size finite-size rounding in
the direction along the wedge 
and the rounding in the two other directions set in simultaneously.\footnote{If we kept the ratio $L_y/L$ constant $\eta \to 0$ and the system would exhibit a 
behavior characteristic of a corner. In the limit $L$ fixed but $L_y$ the wedge becomes quasi-onedimensional and there is no transition.\protect\cite{WEDGE2}}
Hence, the scaling of the probability distribution not only confirms $\beta=0$ but also $\nu_y=3\nu_\perp$.

Knowing the probability distribution of the order parameter we can calculate all its moments:
\begin{equation}
\langle m^k \rangle = {\cal M}_k(L^{1/\nu_\perp}t,\eta)
\label{eqn:moments}
\end{equation}
where ${\cal M}_k$ are scaling functions.
As a special case, we calculate the susceptibility:
$\chi=L^2L_y \langle m^2 \rangle/k_BT \sim L^2L_y \tilde{\cal M}_2(L/\xi_\perp,L_y/\xi_y) \sim \xi_\perp^2\xi_y \sim t^{-2\nu_\perp-\nu_y}\equiv t^{-\gamma}$
with $\gamma = 2\nu_\perp + \nu_y = 5/4$. Gratifyingly these values for the exponents comply with the anisotropic hyperscaling relation\cite{ANI}
$\gamma + 2 \beta =(d-1) \nu_\perp + \nu_y$. Using thermodynamic scaling $2 - \alpha = \gamma + 2 \beta $ we infer the critical exponent
$\alpha= 3/4$ for the specific heat.
Another consequence of the absence of any $L$-dependent prefactor in Eq.(\ref{eqn:moments}) is the common intersection of moments of the
order parameter at the transition. Again this is an agreement with our observation in Fig,\ref{fig:pfirst}({\bf a}). As this intersection 
involves only
the lowest moment of the order parameter it yields an accurate estimate of the location of the critical interface localization/delocalization transition in 
an antisymmetric double wedge.

\begin{figure}[th]
\begin{minipage}{0.5\linewidth}
\epsfig{file=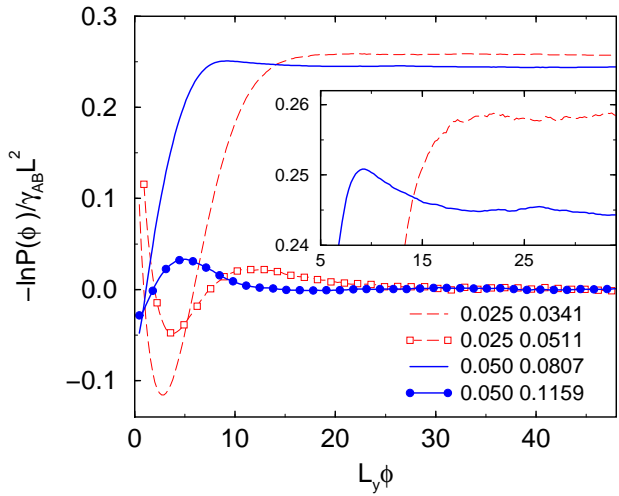,width=\linewidth}
\end{minipage}
\hfill
\begin{minipage}{0.45\linewidth}
\caption{
\label{fig:gspin} 
\protect{
The probability distribution of the composition in an antisymmetric film with system geometry \protect{$48^2\times 96$}.
Values of $\epsilon/k_BT$ and $\epsilon_{\rm wall}/k_BT$ (shown in the key) correspond to the wetting transition and the filling transition (according
to Young's equation).
}
}
\end{minipage}
\end{figure}
It is interesting to relate the observation of first and second order interface localization/delocalization transition in a double wedge
to the shape of the interface potential. Parry {\em et al.\ } predict\cite{PARRYW} that the filling transition is second order if the interface 
potential between an $AB$ interface and a planar surface does not exhibit a free energy barrier between the minimum close to the surface and the 
behavior at large distances, i.e.\ if a macroscopically thick film is not even metastable.

In Fig.\ref{fig:gspin} we present the interface potential obtained from the probability distribution of the composition in a simulation of an antisymmetric film at
$\epsilon/k_BT=0.025$. In the vicinity of the wetting transition the interface potential exhibits a maximum between the minimum close
to the surface and the value at large distances. This fact confirms that the wetting transition is of first order. At the smaller value
of $\epsilon_{\rm wall}$, however, there is no such maximum within the statistical uncertainty of the Monte Carlo data and,
in agreement with Parry's predictions,
we observe a second order transition in the double wedge.

\section{Summary}
We have investigated the interplay between wetting and phase separation of incompressible binary mixtures confined in thin films and wedges.
In our polymer model the wetting transition if of first order and we can accurately locate it via Young's equation. The concomitant pre-wetting
behavior modifies the phase boundaries in thin films. If both surfaces attract the same component, capillary condensation occurs and the critical 
point is close to the critical unmixing transition in the bulk. If one surfaces attracts the $A$-component but the other attracts the $B$-component
an interface localization/delocalization transition occurs. In this case there are two critical points which correspond to the pre-wetting critical
points at each surface. If the film thickness is very small, however, the interface localization/delocalization transition might be of second order
even if the wetting transition is of first order. The critical points in a thin film are characterized by Ising critical behavior.

In analogy to the interface localization/delocalization in an antisymmetric film, we have studied the transition in an antisymmetric double wedge
and we relate the phase behavior to the filling transition in a single wedge. Importantly we present evidence that the analog of critical 
filling in an antisymmetric double wedge geometry gives rise
to unconventional critical behavior characterized by an order parameter exponent $\beta=0$ and strong anisotropic fluctuations\cite{WEDGE}.
We can relate the critical exponents to the predictions of Parry {\em et al.}\cite{PARRYW} on critical filling. In agreement with those
predictions the filling transition can be critical even though the wetting transition on a planar substrate is of first order. This is practically
important because there is no experimental realization of critical wetting on a solid substrate. Our findings suggests the polymer blends might be 
promising candidates to explore the filling behavior experimentally.

%
%
\vspace*{5mm}

{\bf Acknowledgment}:
It is a great pleasure to thank E.V.~Albano, D.P.~Landau and A.~Milchev for fruitful collaborations and A.O.~Parry
for stimulating discussions.
Financial support by the DFG under grants Mu 1674/1-1, Bi314/17-4, and DAAD/PROALAR 2000 as well as
computer time at NIC J{\"u}lich and HLR Stuttgart are acknowledged.
%
%
%
%
%


\end{document}